\documentclass[aps,
twocolumn,unsortedaddress,floatfix]{revtex4-1}
\usepackage{amsthm}
\usepackage{amsfonts}
\usepackage{siunitx}
\usepackage{amsmath}
\usepackage{amssymb}
\usepackage{graphicx}
\usepackage{verbatim}
\usepackage[colorlinks]{hyperref}
\usepackage{tikz}
\usepackage{pgfplots}
\usepackage{braket}
\usepackage{xcolor}

\definecolor{linkcolor}{RGB}{0,83,166}
\hypersetup{
  colorlinks = true,
  allcolors = {linkcolor}
}

\begin{document}

\title{Performance benefits of increased qubit connectivity in quantum annealing 3-dimensional spin glasses}

\author{Andrew D. King and William Bernoudy}

\affiliation{D-Wave Systems, Burnaby, British Columbia, Canada, V5G 4M9, Canada}

\date{\today}

\begin{abstract}An important challenge in superconducting quantum computing is the need to physically couple many devices using quasi-two-dimensional fabrication processes.  Recent advances in the design and fabrication of quantum annealing processors have enabled an increase in pairwise connectivity among thousands of qubits.  One benefit of this is the ability to minor-embed optimization problems using fewer physical qubits for each logical spin.  Here we demonstrate the benefit of this progress in the problem of minimizing the energy of three-dimensional spin glasses.  Comparing the previous generation D-Wave 2000Q system to the new Advantage system, we observe improved scaling of solution time and improved consistency over multiple graph embeddings.
\end{abstract}

\maketitle

\section{Introduction}

Hard computational tasks in fields such as optimization and machine learning can often be expressed as the task of finding or sampling low-energy states from a cost function corresponding to the Hamiltonian of an Ising spin model.  Quantum annealing is a means of performing this task by physically realizing the desired cost function as the terminus of a time-dependent Hamiltonian in the transverse field Ising model (TFIM), in which a transverse magnetic field induces quantum fluctuations, thereby driving dynamics.

The time-dependent TFIM Hamiltonian is
\begin{equation}\label{eq:ham}
\mathcal H(s) = \mathcal J(s) \sum_{\langle ij \rangle}J_{ij}\hat \sigma_i^z \hat\sigma_j^z  - \Gamma(s)\sum_i  \hat \sigma_i^x,
\end{equation}
where the unitless annealing parameter $s$ represents progress through the anneal, $\hat\sigma_i^x$ and $\hat\sigma_i^z$ are Pauli operators, and $\mathcal J(s)$ and $\Gamma(s)$ control the relative magnitude of classical and quantum energy scales \cite{Johnson2011,Harris2018}.

Quantum annealing was motivated in part by adiabatic quantum computation \cite{Farhi2001}, and in part by promising simulations and {\em in situ} experiments on spin glasses \cite{Kadowaki1998,Brooke1999,Santoro2002}. Sampling and optimization in these random bond models are paradigmatic computational challenges.  Three-dimensional spin glasses are of particular interest because they are sparse, physically realistic, computationally intractable, and have a spin-glass phase at finite temperature \cite{Barahona1982,Katzgraber2014, Harris2018}.

In this paper we compare performance of two commercially-available quantum annealing processors on 3D spin glasses: the D-Wave 2000Q LN (first available in 2019), and Advantage (first available in 2020).

\subsection{Chimera versus Pegasus: Old and new qubit connectivity graphs}

Each commercially-available D-Wave processor prior to Advantage had its qubits connected in a {\em Chimera} graph (Fig.~\ref{fig:embedding} top left), consisting of eight-qubit unit cells intercoupled in a square grid.  The designs ranged from 128 to 2048 qubits, some of which are typically (but not always \cite{Harris2018,King2018}) inoperable.

The Advantage system employs a {\em Pegasus} qubit connectivity graph \cite{Boothby2020} (Fig.~\ref{fig:embedding} bottom), which expands on the Chimera graph by increasing connectivity, staggering unit cells, and adding non-bipartite couplers.  In this work we compare a D-Wave 2000Q LN system (2KQ) with 2041 qubits in the Chimera working graph against an Advantage system (ADV) with 5510 qubits in the Pegasus working graph.

\subsection{Embedding 3D lattices}

\begin{figure*}
  \includegraphics[width=.99\linewidth]{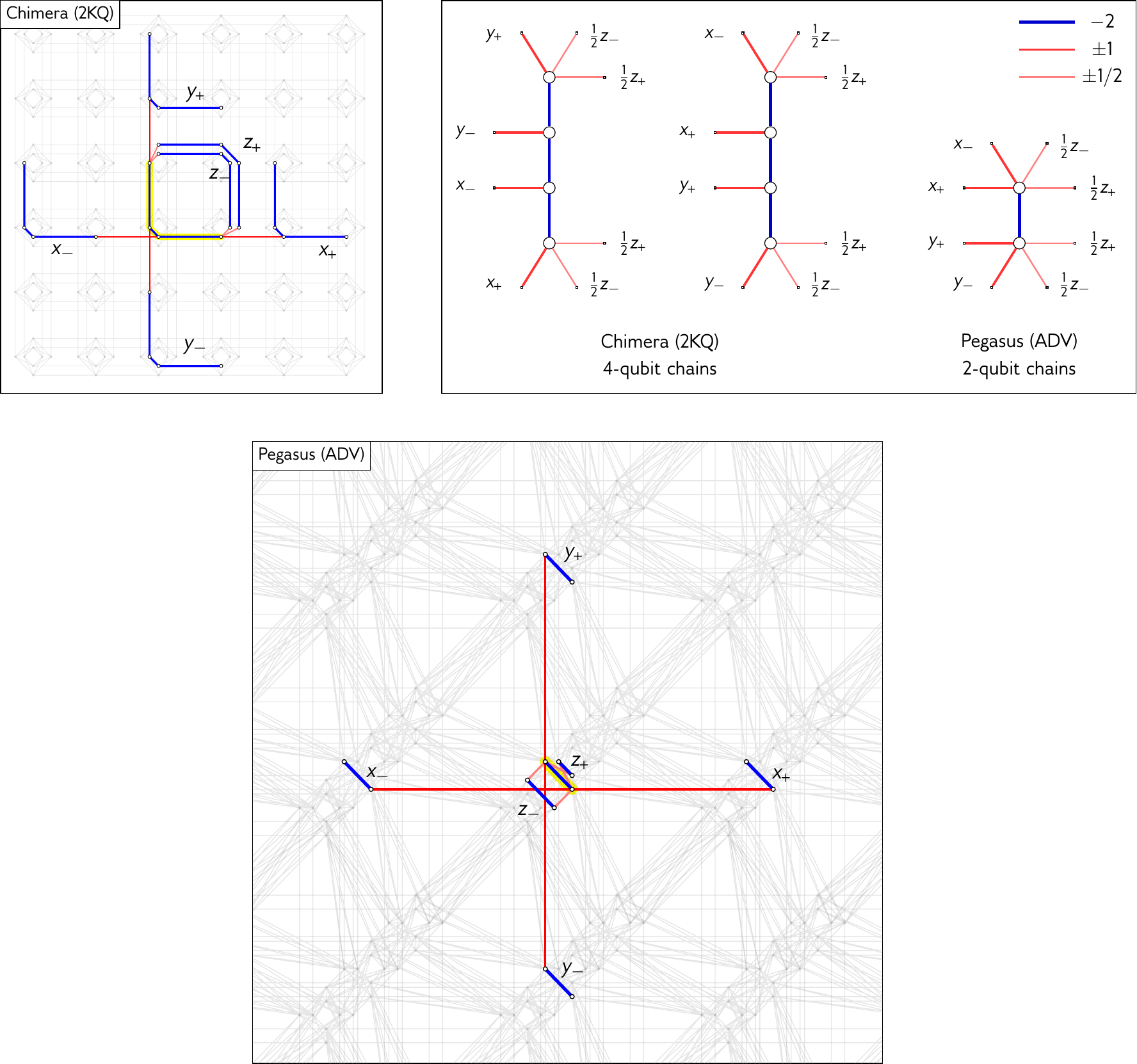}
  \caption{Minor embeddings and chain geometry.  Each logical spin in the 3D spin glass must be represented by a ferromagnetically coupled chain of physical qubits.  Embedding into 2KQ (top left) and ADV (bottom) requires 4- and 2-qubit chains, respectively.  Each logical lattice spin has up to two neighboring lattice spins in each dimension; the corresponding chains are shown for the two embeddings, with the central chain highlighted in yellow.  In 2KQ there are two chain geometries (top right), which alternate along the $z$ dimension (top left).  In ADV there is one unique geometry.  The 2KQ and ADV chains have similar layouts of connections to neighboring spins in the $x$, $y$, and $z$ directions; each logical $z$-coupling is shared across two physical couplers (pale red).}\label{fig:embedding}
\end{figure*}

Since neither 2KQ nor ADV contains a large set of qubits directly coupled in a 3D lattice, these problems must be solved using minor embedding: each logical lattice spin is represented by multiple physical qubits, which are forced to take the same value by strong ferromagnetic couplings connecting them in a chain \cite{Choi2008}.

Ignoring graph defects, a $L_1\times L_2 \times L_3$ lattice can be embedded in 2KQ with four-qubit chains if $L_1$, $L_2$, and $L_3$ are all $\leq 8$.  It can be embedded in ADV with two-qubit chains if $L_1$ and $L_2$ are $\leq 15$, and $L_3 \leq 12$.

The basic geometry of a chain is the same in both embeddings: the $x$-couplings lie on one side of the chain, the $y$-couplings on the other, and each $z$-coupling is divided across two physical couplers, one on each half of the chain (Fig.~\ref{fig:embedding}).  Consequently both embeddings have an ``algebraic chain strength'' of $2$: to guarantee the embedded classical problem always has a ground state with no broken chains, it is necessary and sufficient to set the magnitude of the FM chain couplings to be twice the maximum magnitude of a logical coupling.

In both 2KQ and ADV, the couplings must lie in the range $[-2,1]$.  Therefore setting all chain couplers to $-2$ and logical couplers to $\pm 1$ maximizes both the chain and the logical energy scales \footnote{Since $z$-couplings are split across two physical couplers, each will be given a value of $\pm \tfrac 12$.}.  In the optimization context studied here, this provides approximately optimal performance.

\subsection{Methods}

To compare 2KQ with ADV, we generate embedded 3D spin glass problems of size $L\times L \times L$ for $L$ ranging from $5$ to $10$ with open boundaries, using the same problems for each processor for $L\leq 8$; for $L>8$ the problems are only embeddable in 2KQ.  For this simple study we do not study instances with $L>10$, although they are embeddable in ADV.

For each system size we use the same subgraph of the $L\times L\times L$ lattice, with missing spins and couplers corresponding to inoperable qubits and couplers in the embeddings.  We use the same logical lattice graph for each size and for each processor.  Given a working graph it is possible to maximize the yielded logical graph by making local adjustments to the configurations shown in Fig.~\ref{fig:embedding}, and we find high-yield logical graphs with a heuristic search.  Lattice yield decays with system size: for $L=5,6$ there are no defects; for $L=10$, $972$ of $1000$ logical spins are embedded, using $2550$ of $2700$ logical couplers.  For each $L$ we generate $100$ random instances where each extant coupler is assigned a value $J_{ij}=-1$ or $+1$ uniformly at random.

Problems are run in batches of $500$ anneals of length $t_a = 2, 4, 8,\ldots \SI{256}{\micro s}$ until the presumed ground state has been found at least $50$ times.  Flux-bias offsets are used to balance chains at degeneracy (average magnetization zero) \footnote{This is recommended when using chains; see for example Extended Data Fig.~7 and related methods in \cite{King2018}, and a similar approach in Appendix B of \cite{Nishimura2020}.}.

For a problem with ground state probability $p_{\text{GS}}(t_a)$ and for a given anneal time, we compute the annealing time to solution \cite{Roennow2014}:
\begin{equation}
  \text{TTS}(t_a) = t_a\frac{\log(0.01)}{\log(1-p_{\text{GS}}(t_a)}.
\end{equation}
Of the tested anneal times, we select the optimal anneal time $t_{\text{opt}}$ for an instance as the one that minimizes $p_{\text{GS}}(t_a)$, and select the overall time to solution for the instance as $\text{TTS}=\text{TTS}(t_{\text{opt}})$.

\section{Results}

\subsection{Solution time scaling}

\begin{figure}
  \includegraphics{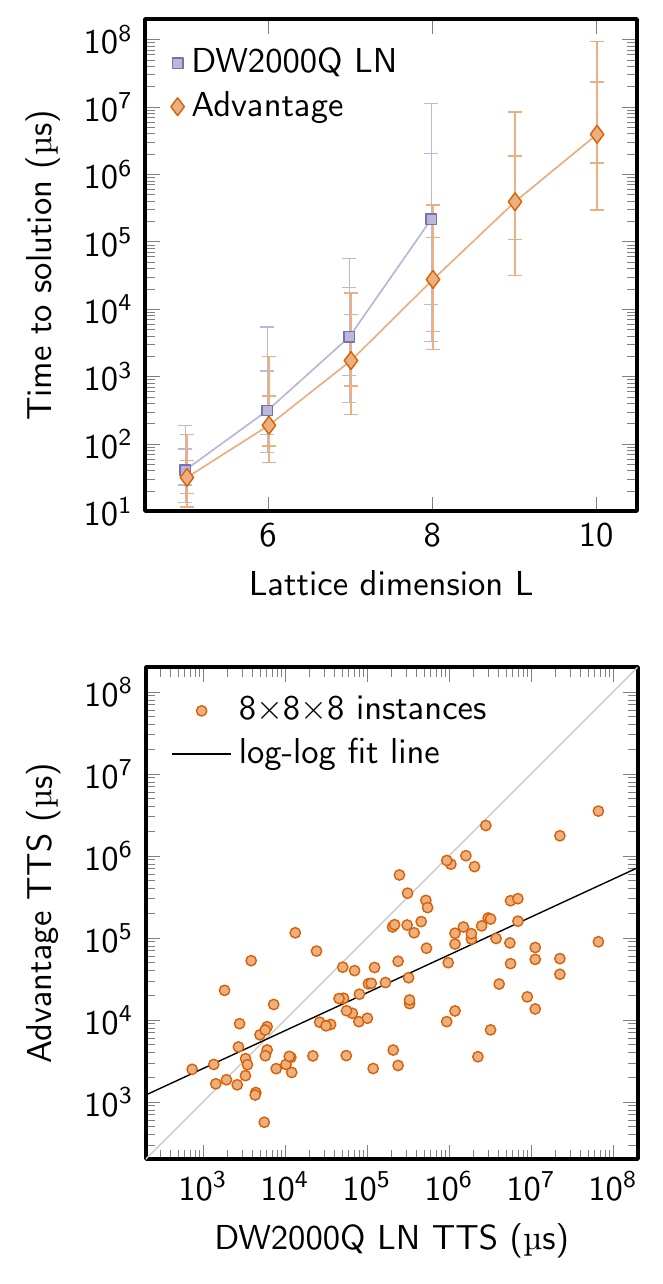}
  \caption{{\bf Top:} Scaling of time to solution.  Data markers indicate median of $100$ instances; error bars indicate percentiles $10$, $25$, $75$, $90$. {\bf Bottom:} Instance-to-instance comparison of $L=8$ instances.  2KQ failed to solve three instances.  On the remaining 97, ADV is up to $800$$\times$ faster.}\label{fig:1}
\end{figure}

\begin{figure*}
  \includegraphics{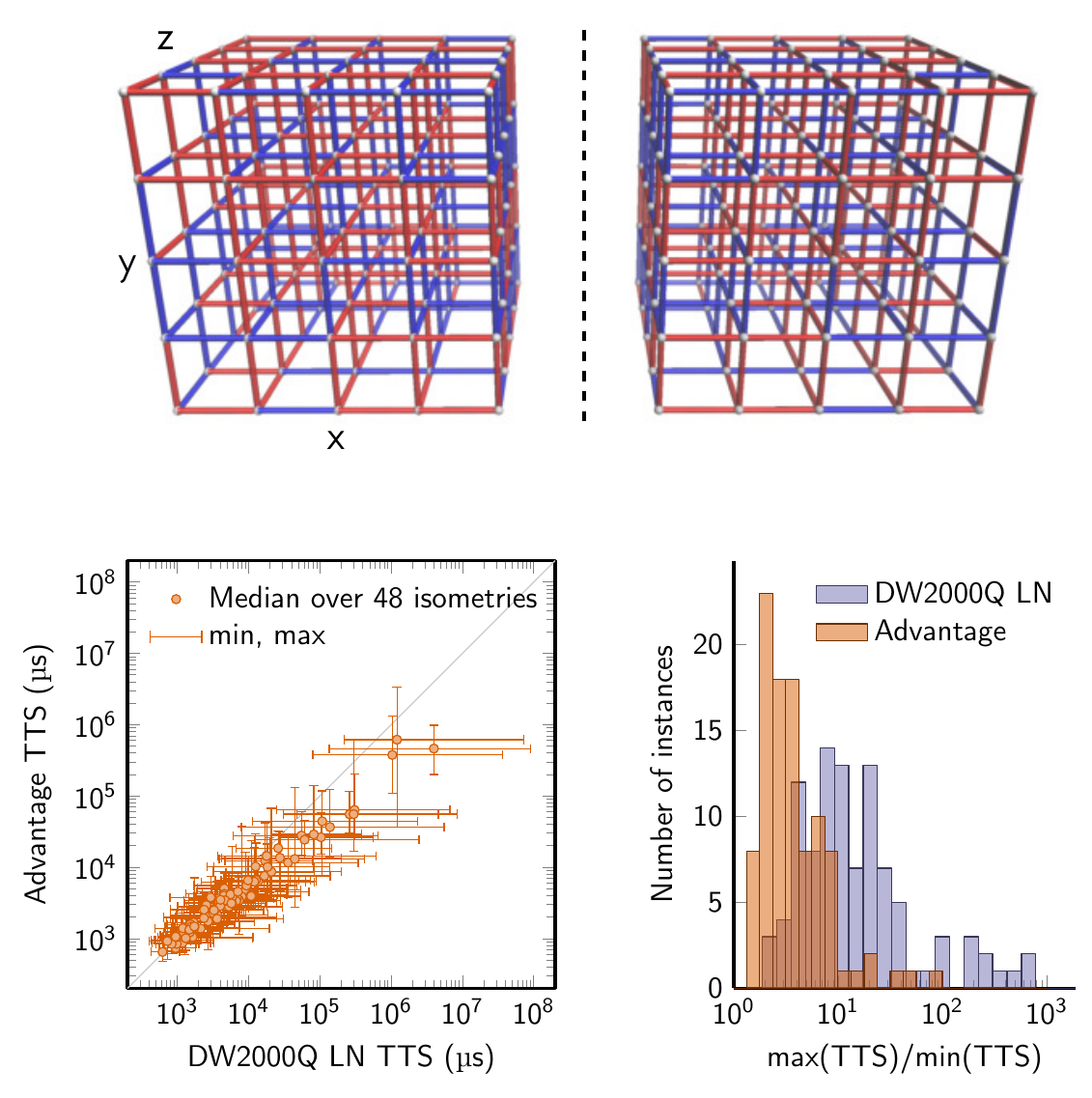}
  \caption{Performance consistency over multiple cube isometries.  {\bf Top:} Two isometries of a $L=5$ spin glass, related by a flip of the $x$ axis.  There are $48$ isometries in total, forming the octahedral symmetry group $O_h$.  {\bf Bottom left:} Instance-to-instance comparison of $L=6$ instances.  Each instance is run on each processor using the 48 cube isometries.  Each data point indicates the median performance across the isometries for the two processors, while the error bars indicate the best and worst performance for individual isometries.  The fact that vertical error bars are smaller than horizontal error bars indicates that performance is more consistent for ADV than for 2KQ.  {\bf Bottom right:} Ratio of best and worst performance (error bar range), histogrammed.}\label{fig:3}
\end{figure*}

Fig.~\ref{fig:1} (top) shows the scaling of time to solution for 2KQ and ADV.  For the systems studied, the results indicate a clear scaling advantage for ADV over 2KQ, in addition to the fact that much larger systems can be run on ADV.

Fig.~\ref{fig:1} (bottom) compares TTS between the two processors for the $L=8$, which is the largest lattice embeddable in 2KQ.  Three of 100 instances are not shown because 2KQ never found the best energy found by ADV.  These results indicate that even within a set of problems of a given size, ADV tends to show the greatest advantage over the 2KQ processor for the hardest problems.  This is a strong indication of systematic performance improvement.

\subsection{Consistency across cube isometries}

For $L=6$, the full 3D lattice can be embedded in both ADV and 2KQ.  The lattices are therefore invariant under the 48 isometries of a cube, giving us 48 embeddings of each problem using the same chains.  To measure distortion in the problem arising from embedding, we compare TTS for 48 isometries of each problem, running only $t_a=\SI{128}{\micro s}$, shown in Fig.~\ref{fig:3}.  In this measurement ADV offers a substantial improvement in consistency compared to 2KQ, which we can attribute primarily to the reduction in chain length.

\section{Conclusions}

Along with the increased connectivity of the Pegasus architecture (from six to fifteen couplers per qubit) come parametric compromises, most notably in energy scale.  For 2KQ the curves $\mathcal J(s)$ and $\Gamma(s)$ cross at an energy scale of $\SI{1.91}{\giga\Hz}$, whereas for ADV they cross at $\SI{1.50}{\giga\Hz}$, $22\%$ lower.  Additionally, 2KQ is based on a production process that has been refined over multiple generations \cite{Chancellor2020}, and has a lower noise profile than ADV, which has been completely redesigned.  Taken along with the improved consistency across cube isometries, this is strong evidence that chain length plays an important role in performance.

Although it is possible to mitigate chain nonidealities to some extent \cite{Raymond2020,Harris2018}, we expect this benefit in chain length to carry over to a multitude of embedded problems.  Many interesting geometries are embeddable with much shorter chains in the Advantage system than in the 2000Q system.  A systematic performance improvement has been demonstrated on cliques and random constraint satisfaction instances \cite{ADVTR1}.
 In frustrated two-dimensional lattices, energy minimization and sampling are relatively easy \cite{Houdayer2001} but quantum simulation is of great interest \cite{Moessner2001,Isakov2003}.  Recent demonstrations on embedded square ice \cite{King2020}, Shastry-Sutherland lattice \cite{Kairys2020}, triangular lattice \cite{King2018,King2019}, and $\mathbb Z_2$ lattice gauge theory \cite{Chamon2019,Zhou2020} all use four-qubit chains.  We expect that recent demonstrations of quantum simulation on these 2D systems can be improved using two qubits instead of four qubits per chain on the Advantage system.

\bibliography{mybib}

\begin{thebibliography}{26}%
\makeatletter
\providecommand \@ifxundefined [1]{%
 \@ifx{#1\undefined}
}%
\providecommand \@ifnum [1]{%
 \ifnum #1\expandafter \@firstoftwo
 \else \expandafter \@secondoftwo
 \fi
}%
\providecommand \@ifx [1]{%
 \ifx #1\expandafter \@firstoftwo
 \else \expandafter \@secondoftwo
 \fi
}%
\providecommand \natexlab [1]{#1}%
\providecommand \enquote  [1]{``#1''}%
\providecommand \bibnamefont  [1]{#1}%
\providecommand \bibfnamefont [1]{#1}%
\providecommand \citenamefont [1]{#1}%
\providecommand \href@noop [0]{\@secondoftwo}%
\providecommand \href [0]{\begingroup \@sanitize@url \@href}%
\providecommand \@href[1]{\@@startlink{#1}\@@href}%
\providecommand \@@href[1]{\endgroup#1\@@endlink}%
\providecommand \@sanitize@url [0]{\catcode `\\12\catcode `\$12\catcode
  `\&12\catcode `\#12\catcode `\^12\catcode `\_12\catcode `\%12\relax}%
\providecommand \@@startlink[1]{}%
\providecommand \@@endlink[0]{}%
\providecommand \url  [0]{\begingroup\@sanitize@url \@url }%
\providecommand \@url [1]{\endgroup\@href {#1}{\urlprefix }}%
\providecommand \urlprefix  [0]{URL }%
\providecommand \Eprint [0]{\href }%
\providecommand \doibase [0]{http://dx.doi.org/}%
\providecommand \selectlanguage [0]{\@gobble}%
\providecommand \bibinfo  [0]{\@secondoftwo}%
\providecommand \bibfield  [0]{\@secondoftwo}%
\providecommand \translation [1]{[#1]}%
\providecommand \BibitemOpen [0]{}%
\providecommand \bibitemStop [0]{}%
\providecommand \bibitemNoStop [0]{.\EOS\space}%
\providecommand \EOS [0]{\spacefactor3000\relax}%
\providecommand \BibitemShut  [1]{\csname bibitem#1\endcsname}%
\let\auto@bib@innerbib\@empty
\bibitem [{\citenamefont {Johnson}\ \emph {et~al.}(2011)\citenamefont
  {Johnson}, \citenamefont {Amin}, \citenamefont {Gildert}, \citenamefont
  {Lanting}, \citenamefont {Hamze} \emph {et~al.}}]{Johnson2011}%
  \BibitemOpen
  \bibfield  {author} {\bibinfo {author} {\bibfnamefont {M.~W.}\ \bibnamefont
  {Johnson}}, \bibinfo {author} {\bibfnamefont {M.~H.}\ \bibnamefont {Amin}},
  \bibinfo {author} {\bibfnamefont {S.}~\bibnamefont {Gildert}}, \bibinfo
  {author} {\bibfnamefont {T.}~\bibnamefont {Lanting}}, \bibinfo {author}
  {\bibfnamefont {F.}~\bibnamefont {Hamze}},  \emph {et~al.},\ }\href {\doibase
  10.1038/nature10012} {\bibfield  {journal} {\bibinfo  {journal} {Nature}\
  }\textbf {\bibinfo {volume} {473}},\ \bibinfo {pages} {194} (\bibinfo {year}
  {2011})}\BibitemShut {NoStop}%
\bibitem [{\citenamefont {Harris}\ \emph {et~al.}(2018)\citenamefont {Harris},
  \citenamefont {Sato}, \citenamefont {Berkley}, \citenamefont {Reis},
  \citenamefont {Altomare} \emph {et~al.}}]{Harris2018}%
  \BibitemOpen
  \bibfield  {author} {\bibinfo {author} {\bibfnamefont {R.}~\bibnamefont
  {Harris}}, \bibinfo {author} {\bibfnamefont {Y.}~\bibnamefont {Sato}},
  \bibinfo {author} {\bibfnamefont {A.~J.}\ \bibnamefont {Berkley}}, \bibinfo
  {author} {\bibfnamefont {M.}~\bibnamefont {Reis}}, \bibinfo {author}
  {\bibfnamefont {F.}~\bibnamefont {Altomare}},  \emph {et~al.},\ }\href
  {\doibase 10.1126/science.aat2025} {\bibfield  {journal} {\bibinfo  {journal}
  {Science}\ }\textbf {\bibinfo {volume} {165}},\ \bibinfo {pages} {162}
  (\bibinfo {year} {2018})}\BibitemShut {NoStop}%
\bibitem [{\citenamefont {Farhi}\ \emph {et~al.}(2001)\citenamefont {Farhi},
  \citenamefont {Goldstone}, \citenamefont {Gutmann}, \citenamefont {Lapan},
  \citenamefont {Lundgren},\ and\ \citenamefont {Preda}}]{Farhi2001}%
  \BibitemOpen
  \bibfield  {author} {\bibinfo {author} {\bibfnamefont {E.}~\bibnamefont
  {Farhi}}, \bibinfo {author} {\bibfnamefont {J.}~\bibnamefont {Goldstone}},
  \bibinfo {author} {\bibfnamefont {S.}~\bibnamefont {Gutmann}}, \bibinfo
  {author} {\bibfnamefont {J.}~\bibnamefont {Lapan}}, \bibinfo {author}
  {\bibfnamefont {A.}~\bibnamefont {Lundgren}}, \ and\ \bibinfo {author}
  {\bibfnamefont {D.}~\bibnamefont {Preda}},\ }\href {\doibase
  10.1126/science.1057726} {\bibfield  {journal} {\bibinfo  {journal}
  {\href{http://www.sciencemag.org/cgi/doi/10.1126/science.1057726}{Science}}\
  }\textbf {\bibinfo {volume} {292}},\ \bibinfo {pages} {472} (\bibinfo {year}
  {2001})}\BibitemShut {NoStop}%
\bibitem [{\citenamefont {Kadowaki}\ and\ \citenamefont
  {Nishimori}(1998)}]{Kadowaki1998}%
  \BibitemOpen
  \bibfield  {author} {\bibinfo {author} {\bibfnamefont {T.}~\bibnamefont
  {Kadowaki}}\ and\ \bibinfo {author} {\bibfnamefont {H.}~\bibnamefont
  {Nishimori}},\ }\href {\doibase 10.1103/PhysRevE.58.5355} {\bibfield
  {journal} {\bibinfo  {journal} {Physical Review E}\ }\textbf {\bibinfo
  {volume} {58}},\ \bibinfo {pages} {5355} (\bibinfo {year}
  {1998})}\BibitemShut {NoStop}%
\bibitem [{\citenamefont {Brooke}\ \emph {et~al.}(1999)\citenamefont {Brooke},
  \citenamefont {Bitko}, \citenamefont {F.}, \citenamefont {Rosenbaum},\ and\
  \citenamefont {Aeppli}}]{Brooke1999}%
  \BibitemOpen
  \bibfield  {author} {\bibinfo {author} {\bibfnamefont {J.}~\bibnamefont
  {Brooke}}, \bibinfo {author} {\bibfnamefont {D.}~\bibnamefont {Bitko}},
  \bibinfo {author} {\bibfnamefont {T.}~\bibnamefont {F.}}, \bibinfo {author}
  {\bibnamefont {Rosenbaum}}, \ and\ \bibinfo {author} {\bibfnamefont
  {G.}~\bibnamefont {Aeppli}},\ }\href {\doibase 10.1126/science.284.5415.779}
  {\bibfield  {journal} {\bibinfo  {journal} {Science}\ }\textbf {\bibinfo
  {volume} {284}},\ \bibinfo {pages} {779 } (\bibinfo {year}
  {1999})}\BibitemShut {NoStop}%
\bibitem [{\citenamefont {Santoro}\ \emph {et~al.}(2002)\citenamefont
  {Santoro}, \citenamefont {Marton{\'{a}}k}, \citenamefont {Tosatti},\ and\
  \citenamefont {Car}}]{Santoro2002}%
  \BibitemOpen
  \bibfield  {author} {\bibinfo {author} {\bibfnamefont {G.~E.}\ \bibnamefont
  {Santoro}}, \bibinfo {author} {\bibfnamefont {R.}~\bibnamefont
  {Marton{\'{a}}k}}, \bibinfo {author} {\bibfnamefont {E.}~\bibnamefont
  {Tosatti}}, \ and\ \bibinfo {author} {\bibfnamefont {R.}~\bibnamefont
  {Car}},\ }\href {\doibase 10.1126/science.1068774} {\bibfield  {journal}
  {\bibinfo  {journal} {Science}\ }\textbf {\bibinfo {volume} {295}},\ \bibinfo
  {pages} {2427} (\bibinfo {year} {2002})}\BibitemShut {NoStop}%
\bibitem [{\citenamefont {Barahona}(1982)}]{Barahona1982}%
  \BibitemOpen
  \bibfield  {author} {\bibinfo {author} {\bibfnamefont {F.}~\bibnamefont
  {Barahona}},\ }\href {\doibase 10.1088/0305-4470/15/10/028} {\bibfield
  {journal} {\bibinfo  {journal} {Journal of Physics A: Mathematical and
  General}\ }\textbf {\bibinfo {volume} {15}},\ \bibinfo {pages} {3241}
  (\bibinfo {year} {1982})}\BibitemShut {NoStop}%
\bibitem [{\citenamefont {Katzgraber}\ \emph {et~al.}(2014)\citenamefont
  {Katzgraber}, \citenamefont {Hamze},\ and\ \citenamefont
  {Andrist}}]{Katzgraber2014}%
  \BibitemOpen
  \bibfield  {author} {\bibinfo {author} {\bibfnamefont {H.~G.}\ \bibnamefont
  {Katzgraber}}, \bibinfo {author} {\bibfnamefont {F.}~\bibnamefont {Hamze}}, \
  and\ \bibinfo {author} {\bibfnamefont {R.~S.}\ \bibnamefont {Andrist}},\
  }\href {\doibase 10.1103/PhysRevX.4.021008} {\bibfield  {journal} {\bibinfo
  {journal} {Physical Review X}\ }\textbf {\bibinfo {volume} {4}},\ \bibinfo
  {pages} {021008} (\bibinfo {year} {2014})}\BibitemShut {NoStop}%
\bibitem [{\citenamefont {King}\ \emph {et~al.}(2018)\citenamefont {King},
  \citenamefont {Carrasquilla}, \citenamefont {Raymond}, \citenamefont
  {Ozfidan}, \citenamefont {Andriyash} \emph {et~al.}}]{King2018}%
  \BibitemOpen
  \bibfield  {author} {\bibinfo {author} {\bibfnamefont {A.~D.}\ \bibnamefont
  {King}}, \bibinfo {author} {\bibfnamefont {J.}~\bibnamefont {Carrasquilla}},
  \bibinfo {author} {\bibfnamefont {J.}~\bibnamefont {Raymond}}, \bibinfo
  {author} {\bibfnamefont {I.}~\bibnamefont {Ozfidan}}, \bibinfo {author}
  {\bibfnamefont {E.}~\bibnamefont {Andriyash}},  \emph {et~al.},\ }\href
  {\doibase 10.1038/s41586-018-0410-x} {\bibfield  {journal} {\bibinfo
  {journal} {Nature}\ }\textbf {\bibinfo {volume} {560}},\ \bibinfo {pages}
  {456} (\bibinfo {year} {2018})}\BibitemShut {NoStop}%
\bibitem [{\citenamefont {Boothby}\ \emph {et~al.}(2020)\citenamefont
  {Boothby}, \citenamefont {Bunyk}, \citenamefont {Raymond},\ and\
  \citenamefont {Roy}}]{Boothby2020}%
  \BibitemOpen
  \bibfield  {author} {\bibinfo {author} {\bibfnamefont {K.}~\bibnamefont
  {Boothby}}, \bibinfo {author} {\bibfnamefont {P.}~\bibnamefont {Bunyk}},
  \bibinfo {author} {\bibfnamefont {J.}~\bibnamefont {Raymond}}, \ and\
  \bibinfo {author} {\bibfnamefont {A.}~\bibnamefont {Roy}},\ }\href
  {http://arxiv.org/abs/2003.00133} {\enquote {\bibinfo {title}
  {{Next-Generation Topology of D-Wave Quantum Processors}},}\ } (\bibinfo
  {year} {2020}),\ \Eprint {http://arxiv.org/abs/2003.00133} {arXiv:2003.00133}
  \BibitemShut {NoStop}%
\bibitem [{\citenamefont {Choi}(2008)}]{Choi2008}%
  \BibitemOpen
  \bibfield  {author} {\bibinfo {author} {\bibfnamefont {V.}~\bibnamefont
  {Choi}},\ }\href {\doibase 10.1007/s11128-008-0082-9} {\bibfield  {journal}
  {\bibinfo  {journal} {Quantum Information Processing}\ }\textbf {\bibinfo
  {volume} {7}},\ \bibinfo {pages} {193} (\bibinfo {year} {2008})}\BibitemShut
  {NoStop}%
\bibitem [{Note1()}]{Note1}%
  \BibitemOpen
  \bibinfo {note} {Since $z$-couplings are split across two physical couplers,
  each will be given a value of $\pm \protect \genfrac {}{}{}112$.}\BibitemShut
  {Stop}%
\bibitem [{Note2()}]{Note2}%
  \BibitemOpen
  \bibinfo {note} {This is recommended when using chains; see for example
  Extended Data Fig.~7 and related methods in \cite {King2018}, and a similar
  approach in Appendix B of \cite {Nishimura2020}.}\BibitemShut {Stop}%
\bibitem [{\citenamefont {R{\o}nnow}\ \emph {et~al.}(2014)\citenamefont
  {R{\o}nnow}, \citenamefont {Wang}, \citenamefont {Job}, \citenamefont
  {Boixo}, \citenamefont {Isakov}, \citenamefont {Wecker}, \citenamefont
  {Martinis}, \citenamefont {Lidar},\ and\ \citenamefont
  {Troyer}}]{Roennow2014}%
  \BibitemOpen
  \bibfield  {author} {\bibinfo {author} {\bibfnamefont {T.~F.}\ \bibnamefont
  {R{\o}nnow}}, \bibinfo {author} {\bibfnamefont {Z.}~\bibnamefont {Wang}},
  \bibinfo {author} {\bibfnamefont {J.}~\bibnamefont {Job}}, \bibinfo {author}
  {\bibfnamefont {S.}~\bibnamefont {Boixo}}, \bibinfo {author} {\bibfnamefont
  {S.~V.}\ \bibnamefont {Isakov}}, \bibinfo {author} {\bibfnamefont
  {D.}~\bibnamefont {Wecker}}, \bibinfo {author} {\bibfnamefont {J.~M.}\
  \bibnamefont {Martinis}}, \bibinfo {author} {\bibfnamefont {D.~A.}\
  \bibnamefont {Lidar}}, \ and\ \bibinfo {author} {\bibfnamefont
  {M.}~\bibnamefont {Troyer}},\ }\href {\doibase 10.1126/science.1252319}
  {\bibfield  {journal} {\bibinfo  {journal} {Science}\ }\textbf {\bibinfo
  {volume} {345}},\ \bibinfo {pages} {420} (\bibinfo {year}
  {2014})}\BibitemShut {NoStop}%
\bibitem [{\citenamefont {Chancellor}\ and\ \citenamefont
  {Kendon}(2020)}]{Chancellor2020}%
  \BibitemOpen
  \bibfield  {author} {\bibinfo {author} {\bibfnamefont {N.}~\bibnamefont
  {Chancellor}}\ and\ \bibinfo {author} {\bibfnamefont {V.}~\bibnamefont
  {Kendon}},\ }\href {http://arxiv.org/abs/2008.11054} {\enquote {\bibinfo
  {title} {{Search range in experimental quantum annealing}},}\ } (\bibinfo
  {year} {2020}),\ \Eprint {http://arxiv.org/abs/2008.11054} {arXiv:2008.11054}
  \BibitemShut {NoStop}%
\bibitem [{\citenamefont {Raymond}\ \emph {et~al.}(2020)\citenamefont
  {Raymond}, \citenamefont {Ndiaye}, \citenamefont {Rayaprolu},\ and\
  \citenamefont {King}}]{Raymond2020}%
  \BibitemOpen
  \bibfield  {author} {\bibinfo {author} {\bibfnamefont {J.}~\bibnamefont
  {Raymond}}, \bibinfo {author} {\bibfnamefont {N.}~\bibnamefont {Ndiaye}},
  \bibinfo {author} {\bibfnamefont {G.}~\bibnamefont {Rayaprolu}}, \ and\
  \bibinfo {author} {\bibfnamefont {A.}~\bibnamefont {King}},\ }\href
  {http://arxiv.org/abs/2006.04913} {\enquote {\bibinfo {title} {{Improving
  performance of logical qubits by parameter tuning and topology
  compensation}},}\ } (\bibinfo {year} {2020}),\ \Eprint
  {http://arxiv.org/abs/2006.04913} {arXiv:2006.04913} \BibitemShut {NoStop}%
\bibitem [{\citenamefont {{McGeoch}}\ and\ \citenamefont
  {Farr\'{e}}(2020)}]{ADVTR1}%
  \BibitemOpen
  \bibfield  {author} {\bibinfo {author} {\bibfnamefont {C.}~\bibnamefont
  {{McGeoch}}}\ and\ \bibinfo {author} {\bibfnamefont {P.}~\bibnamefont
  {Farr\'{e}}},\ }\href@noop {} {\enquote {\bibinfo {title}
  {\href{http://dwavesys.com/resources/publications?type=white}{The D-Wave
  Advantage System: An Overview}},}\ } (\bibinfo {year} {2020})\BibitemShut
  {NoStop}%
\bibitem [{\citenamefont {Houdayer}(2001)}]{Houdayer2001}%
  \BibitemOpen
  \bibfield  {author} {\bibinfo {author} {\bibfnamefont {J.}~\bibnamefont
  {Houdayer}},\ }\href {\doibase 10.1007/PL00011151} {\bibfield  {journal}
  {\bibinfo  {journal} {The European Physical Journal B}\ }\textbf {\bibinfo
  {volume} {22}},\ \bibinfo {pages} {479} (\bibinfo {year} {2001})}\BibitemShut
  {NoStop}%
\bibitem [{\citenamefont {Moessner}\ and\ \citenamefont
  {Sondhi}(2001)}]{Moessner2001}%
  \BibitemOpen
  \bibfield  {author} {\bibinfo {author} {\bibfnamefont {R.}~\bibnamefont
  {Moessner}}\ and\ \bibinfo {author} {\bibfnamefont {S.~L.}\ \bibnamefont
  {Sondhi}},\ }\href {\doibase 10.1103/PhysRevB.63.224401} {\bibfield
  {journal} {\bibinfo  {journal} {Physical Review B}\ }\textbf {\bibinfo
  {volume} {63}},\ \bibinfo {pages} {1} (\bibinfo {year} {2001})}\BibitemShut
  {NoStop}%
\bibitem [{\citenamefont {Isakov}\ and\ \citenamefont
  {Moessner}(2003)}]{Isakov2003}%
  \BibitemOpen
  \bibfield  {author} {\bibinfo {author} {\bibfnamefont {S.~V.}\ \bibnamefont
  {Isakov}}\ and\ \bibinfo {author} {\bibfnamefont {R.}~\bibnamefont
  {Moessner}},\ }\href {\doibase 10.1103/PhysRevB.68.104409} {\bibfield
  {journal} {\bibinfo  {journal} {Physical Review B}\ }\textbf {\bibinfo
  {volume} {68}},\ \bibinfo {pages} {104409} (\bibinfo {year}
  {2003})}\BibitemShut {NoStop}%
\bibitem [{\citenamefont {King}\ \emph {et~al.}(2020)\citenamefont {King},
  \citenamefont {Nisoli}, \citenamefont {Dahl}, \citenamefont
  {{Poulin-Lamarre}},\ and\ \citenamefont {{Lopez-Bezanilla}}}]{King2020}%
  \BibitemOpen
  \bibfield  {author} {\bibinfo {author} {\bibfnamefont {A.~D.}\ \bibnamefont
  {King}}, \bibinfo {author} {\bibfnamefont {C.}~\bibnamefont {Nisoli}},
  \bibinfo {author} {\bibfnamefont {E.~D.}\ \bibnamefont {Dahl}}, \bibinfo
  {author} {\bibfnamefont {G.}~\bibnamefont {{Poulin-Lamarre}}}, \ and\
  \bibinfo {author} {\bibfnamefont {A.}~\bibnamefont {{Lopez-Bezanilla}}},\
  }\href {http://arxiv.org/abs/2007.10555} {\enquote {\bibinfo {title}
  {{Quantum Artificial Spin Ice}},}\ } (\bibinfo {year} {2020}),\ \Eprint
  {http://arxiv.org/abs/arXiv:2007.10555} {arXiv:2007.10555} \BibitemShut
  {NoStop}%
\bibitem [{\citenamefont {Kairys}\ \emph {et~al.}(2020)\citenamefont {Kairys},
  \citenamefont {King}, \citenamefont {Ozfidan}, \citenamefont {Boothby},
  \citenamefont {Raymond}, \citenamefont {Banerjee},\ and\ \citenamefont
  {Humble}}]{Kairys2020}%
  \BibitemOpen
  \bibfield  {author} {\bibinfo {author} {\bibfnamefont {P.}~\bibnamefont
  {Kairys}}, \bibinfo {author} {\bibfnamefont {A.~D.}\ \bibnamefont {King}},
  \bibinfo {author} {\bibfnamefont {I.}~\bibnamefont {Ozfidan}}, \bibinfo
  {author} {\bibfnamefont {K.}~\bibnamefont {Boothby}}, \bibinfo {author}
  {\bibfnamefont {J.}~\bibnamefont {Raymond}}, \bibinfo {author} {\bibfnamefont
  {A.}~\bibnamefont {Banerjee}}, \ and\ \bibinfo {author} {\bibfnamefont
  {T.~S.}\ \bibnamefont {Humble}},\ }\href {http://arxiv.org/abs/2003.01019}
  {\enquote {\bibinfo {title} {{Simulating the Shastry-Sutherland Ising Model
  using Quantum Annealing}},}\ } (\bibinfo {year} {2020}),\ \Eprint
  {http://arxiv.org/abs/arXiv:2003.01019} {arXiv:2003.01019} \BibitemShut
  {NoStop}%
\bibitem [{\citenamefont {King}\ \emph {et~al.}(2019)\citenamefont {King},
  \citenamefont {Raymond}, \citenamefont {Lanting}, \citenamefont {Isakov},
  \citenamefont {Mohseni} \emph {et~al.}}]{King2019}%
  \BibitemOpen
  \bibfield  {author} {\bibinfo {author} {\bibfnamefont {A.~D.}\ \bibnamefont
  {King}}, \bibinfo {author} {\bibfnamefont {J.}~\bibnamefont {Raymond}},
  \bibinfo {author} {\bibfnamefont {T.}~\bibnamefont {Lanting}}, \bibinfo
  {author} {\bibfnamefont {S.~V.}\ \bibnamefont {Isakov}}, \bibinfo {author}
  {\bibfnamefont {M.}~\bibnamefont {Mohseni}},  \emph {et~al.},\ }\href
  {http://arxiv.org/abs/1911.03446} {\enquote {\bibinfo {title} {{Scaling
  advantage in quantum simulation of geometrically frustrated magnets}},}\ }
  (\bibinfo {year} {2019}),\ \Eprint {http://arxiv.org/abs/arXiv:1911.03446}
  {arXiv:1911.03446} \BibitemShut {NoStop}%
\bibitem [{\citenamefont {Chamon}\ \emph {et~al.}(2020)\citenamefont {Chamon},
  \citenamefont {Green},\ and\ \citenamefont {Yang}}]{Chamon2019}%
  \BibitemOpen
  \bibfield  {author} {\bibinfo {author} {\bibfnamefont {C.}~\bibnamefont
  {Chamon}}, \bibinfo {author} {\bibfnamefont {D.}~\bibnamefont {Green}}, \
  and\ \bibinfo {author} {\bibfnamefont {Z.-C.}\ \bibnamefont {Yang}},\ }\href
  {\doibase 10.1103/PhysRevLett.125.067203} {\bibfield  {journal} {\bibinfo
  {journal} {Physical Review Letters}\ }\textbf {\bibinfo {volume} {125}},\
  \bibinfo {pages} {067203} (\bibinfo {year} {2020})}\BibitemShut {NoStop}%
\bibitem [{\citenamefont {Zhou}\ \emph {et~al.}(2020)\citenamefont {Zhou},
  \citenamefont {Green}, \citenamefont {Dahl},\ and\ \citenamefont
  {Chamon}}]{Zhou2020}%
  \BibitemOpen
  \bibfield  {author} {\bibinfo {author} {\bibfnamefont {S.}~\bibnamefont
  {Zhou}}, \bibinfo {author} {\bibfnamefont {D.}~\bibnamefont {Green}},
  \bibinfo {author} {\bibfnamefont {E.~D.}\ \bibnamefont {Dahl}}, \ and\
  \bibinfo {author} {\bibfnamefont {C.}~\bibnamefont {Chamon}},\ }\href
  {http://arxiv.org/abs/2009.07853} {\enquote {\bibinfo {title} {{Building and
  Probing Spin Liquids in a Programmable Quantum Device}},}\ } (\bibinfo {year}
  {2020}),\ \Eprint {http://arxiv.org/abs/2009.07853} {arXiv:2009.07853}
  \BibitemShut {NoStop}%
\bibitem [{\citenamefont {Nishimura}\ \emph {et~al.}(2020)\citenamefont
  {Nishimura}, \citenamefont {Nishimori},\ and\ \citenamefont
  {Katzgraber}}]{Nishimura2020}%
  \BibitemOpen
  \bibfield  {author} {\bibinfo {author} {\bibfnamefont {K.}~\bibnamefont
  {Nishimura}}, \bibinfo {author} {\bibfnamefont {H.}~\bibnamefont
  {Nishimori}}, \ and\ \bibinfo {author} {\bibfnamefont {H.~G.}\ \bibnamefont
  {Katzgraber}},\ }\href {http://arxiv.org/abs/2006.16219} {\enquote {\bibinfo
  {title} {{Griffiths-McCoy singularity on the diluted Chimera graph: Monte
  Carlo simulations and experiments on the quantum hardware}},}\ } (\bibinfo
  {year} {2020}),\ \Eprint {http://arxiv.org/abs/2006.16219} {arXiv:2006.16219}
  \BibitemShut {NoStop}%
\end{thebibliography}%

\end{document}